\documentstyle[12pt]{article}
\newcommand{\be}{\begin{eqnarray}}
\newcommand{\ee}{\end{eqnarray}}

\input{epsfig.sty}

\begin{document}

\begin{center}

{\Large Quasi-elastic and inelastic inclusive electron scattering
from an oxygen jet target \footnote{\bf{To appear in Nucl. Phys. A.}}}

\vspace{2cm}

{\large M. Anghinolfi, M. Ripani, M. Battaglieri, R. Cenni, P.
Corvisiero,\\ A. Longhi, V.I. Mokeev, G. Ricco, M. Taiuti, A. Teglia,
A. Zucchiatti}

\vspace{0.25cm}

{\normalsize{\em Physics Department, University of Genova and
Istituto Nazionale di Fisica Nucleare -\\ Sezione di Genova
Via Dodecaneso 33, I-16146 Genova (Italy)}}

\vspace{1cm}

{\large N. Bianchi, A. Fantoni, P. Levi Sandri, V. Lucherini, V.
Muccifora,\\ E. Polli, A. Reolon, P. Rossi}

\vspace{0.25cm}

{\normalsize{\em Istituto Nazionale di Fisica Nucleare - Laboratori
Nazionali di Frascati\\ P. O. Box 13, I-00044 Frascati (Italy)}}

\vspace{1cm}

{\large S. Simula}

\vspace{0.25cm}

{\normalsize{\em Istituto Nazionale di Fisica Nucleare - Sezione
Sanit\`a,\\ Viale Regina Elena 299, I-00161 Rome (Italy)}}

\vspace{2cm}

\begin{abstract}
The results of an experiment on inclusive electron scattering from an
oxygen jet target, performed in a wide range of energy and momentum
transfer covering both quasi-elastic and $\Delta$(1232) resonance
regions, are reported. In the former region the theoretical predictions,
obtained including effects of nucleon-nucleon correlations in both
initial and final states, give a good description of the experimental
data. In the inelastic region a broadening as well as a  damping of the
resonant part of the cross section with respect to the free nucleon case
is  observed. The need of more detailed calculations including nuclear
structure effects on the electroproduction cross section of nucleon
resonances is highlighted.
\end{abstract}

\end{center}

\vspace{1cm}

\noindent PACS 25.30.F  ---  Keywords: inclusive electron scattering,
nuclear medium effects.

\newpage

\pagestyle{plain}

\section{Introduction}

Investigation of inclusive electron scattering processes $A(e,e')X$ off
nuclei at high momentum transfer can provide relevant information on
the nuclear wave function and, at values of energy transfer above pion
production threshold, on excitation, propagation and decay of nucleon
resonances in nuclear medium. To this purpose several experiments have 
been performed [1--4] showing that, at values of squared four-momentum
transfer $Q^2\sim 0.1 - 1$ (GeV/c)$^2$, the inclusive cross section, as a
function of the energy transfer $\omega$, is characterized by two broad
and prominent peaks which are clearly related to the processes of 
quasi-elastic (QE) scattering and $\Delta$(1232) resonance
electroproduction. As a matter of fact, the centroids of the two peaks
are approximately located at $W \cong M  = 938$ MeV and  $W \cong
M_{\Delta}  = 1232$ MeV respectively [5], where $W \equiv \sqrt{M^2 +
Q^2(1/x - 1)}$ is the invariant mass  produced on a free nucleon at rest
and $x = Q^2 / 2M\omega$ is the Bjorken scaling variable. Thus, the
general features of the inclusive cross section for the $A(e,e')X$ 
reaction are expected to be dominated by the virtual photon absorption
on a quasi-free nucleon. However, such a simple  picture holds only for
kinematical conditions close to the centre of the QE peak (i.e., at 
$x \sim 1$), where the overall behaviour of the total cross section can be
accounted for by calculations based on the plane wave Impulse
Approximation (IA) using a mean-field description of the nuclear
structure [6]. It should be pointed out that reaction mechanisms 
different from the quasi-free one can contribute to the total cross
section at kinematics  corresponding to both sides of the QE peak. In the
low energy side ($x > 1$) the inclusive  cross section is sensitive both
to nuclear binding effects (i.e., to high momentum and high  removal
energy components generated in the nuclear wave function by
nucleon-nucleon (NN) short-range and tensor correlations) and to Final
State Interaction (FSI) effects between the knocked-out nucleon and the
residual nuclear system (see for example ref. [7]). In the high energy
side of QE peak ($x < 1$) contributions to the total cross section
arising from non-nucleonic degrees of freedom and inelastic nucleonic
channels become, in  addition, relevant. In kinematical regions
corresponding to the $\Delta$(1232) resonance excitation,  the picture
that emerges from existing experimental data [1--4] can be summarized as
follows: {\em i)\/} both width and height as well as location of the
$\Delta$(1232) peak are modified by medium  effects but, at the same
time, the total cross section per nucleon scales with the mass number 
$A$; {\em ii)\/} the inclusive cross sections measured in the dip region
at low $Q^2$ ($Q^2 \leq 0.2$ (GeV/c)$^2$)  between the QE and
$\Delta$(1232) peaks, are higher than the theoretical prediction, even
when the  effects resulting from pion production and the corrections due
to Meson Exchange Currents  (MEC) are included [8, 9]. Medium effects,
such as Fermi motion, nuclear binding, Pauli blocking and pion
reabsorption, are thought to be responsible of the modifications of the 
width of $\Delta$(1232) resonance and of its location in energy. It
should also be pointed out that in the case of real photons the
excitation of nucleon resonances with masses above the $\Delta$(1232)
seems to be sizably suppressed in nuclei, leading to a damping of such
resonances from the total nuclear photoabsorption cross section in this
region [10]. Thus, the use of virtual photons to investigate the
excitation of nucleon resonances in nuclei could be of great relevance,
providing information on how baryon structure is affected by the presence
of other nucleons. In brief, measurement of inclusive cross section for
$A(e,e')X$ processes at intermediate values of $Q^2$ ($\sim 0.1 - 1$
(GeV/c)$^2$) still represents a powerful tool to investigate  both the
nuclear structure and medium-dependent modifications of electroexcitation
of the most prominent nucleon resonances.

The aim of this paper is to report on an inclusive electron scattering
experiment  performed at ADONE storage ring at Frascati using an oxygen
jet target and a shower calorimeter. The apparatus allowed the
simultaneous measurement of inclusive cross section in a wide range of
values of energy transfer, ranging from quasi-elastic peak to
kinematical  regions beyond the $\Delta$(1232) resonance, at values of
three-momentum transfer $q \equiv |\mbox{\bf q}|$ up to $\sim
800$ MeV/c. This paper is organized as follows. The experimental
apparatus is described in section 2. The data analysis, pair production
subtraction and radiative corrections are  discussed in detail in
section 3. A comparison of the experimental data with theoretical 
predictions both in QE and in $\Delta$(1232) resonance regions is
presented in section 4. The main conclusions are summarized in section 5.

\section{Experimental apparatus}

The experiment was performed at ADONE, the Frascati storage ring, using
0.5 through 1.5 GeV electrons scattered from a clustered jet target [11]
placed on a straight section of the ring. At the interaction point the
jet was a 6 mm spot whereas the electron beam had a  dimension of 3 mm
FWHM. At each run the stored current was initially 50 mA, the beam 
lifetime was 40 minutes for the typical $\sim 1$ ng/cm$^2$ target
density and the luminosity varied from$10^{-31}$ cm$^{-2}$ s$^{-1}$ down
to a factor 5 lower at the end of each measurement. Due to this
exponential decrease an on-line monitor of the luminosity was necessary
and a detection of M\o ller electrons was used to this purpose, as
described later on. The scattered electrons were  measured at 32$^0$,
37.1$^0$, 83$^0$ at different energies from 0.5 to 1.5 GeV with a
scintillation  detector composed by a front part which allowed mass
separation and angular definition and  by a rear part consisting of BGO
crystals to measure the energy of the scattered electrons and  to improve
their separation from proton and pion background [12]. Despite the common
use  of magnetic spectrometers in electron scattering experiments, our
calorimeter provided a  valid alternative for this apparatus since an
energy resolution of few percent was required,  still sufficient to
separate the broad structures of QE and $\Delta$(1232) peaks. Different 
components were present in this detector:

\begin{enumerate}

\item a telescopic system of small plastic scintillators to discriminate
against neutral particles and to define the solid angle, which could be
varied from 4 to 30 msr depending on the position of the detector
(forward or backward angles respectively);

\item an aerogel Cherenkov detector with a refraction index $n = 1.045$ to
separate electrons from pions with momentum lower than 0.5 GeV/c and
protons. The counter efficiency was measured [13] using electron pairs
produced by the Frascati tagged photon  beam incident on a radiator and
it turned out to be 97\%;

\item a BGO pre-shower of 2.5 cm thickness to improve the separation of
electrons from charged heavier particles: as a matter of fact, the
energy deposited in this device is markedly different for the radiating
electron with respect to an ionizing heavier particle;

\item the shower calorimeter consisting of 20 BGO crystals of 24 radiation
lengths  thickness contained in a carbon fiber housing which is part of a
$4\pi$ electromagnetic calorimeter [14]. Besides giving the sum of the
energies deposited in each crystal, the granularity of this detector was
used as a further test to distinguish electrons from the other particles.

\end{enumerate}

The detector temperature was monitored by several thermocouples and kept
constant  by a temperature control system, whereas LEDs were used to
monitor possible PMT gain variations. Proton events as measured in the
pre-shower ($dE/dx$) and in the calorimeter ($E$)  produced a narrow
line [15], providing a precise complementary method to monitor such 
fluctuations, always limited within 3\%\ and corrected in the off-line
analysis.

Due to this multiparametric information a rejection better than~99\%\ of
hadrons against  electrons was obtained, the electron total
identification efficiency being 95\%\ in all our energy range. This
experimental equipment was also designed to achieve a complete
electromagnetic shower absorption in the calorimeter and therefore no
response function unfolding procedure had to be applied to the collected
data. The energy of scattered electrons was determined as the sum of the
energies released in the calorimeter and in the preshower, the energy
loss in the thin plastics being negligible. The final energy resolution
was moderate ($\sim 2.5$\%\ FWHM  for 1.5 GeV electrons) but definitely
sufficient to separate QE and $\Delta$(1232) peaks.

A ($dE/dx-E$) plastic telescope was used to detect the monokinetic M\o
ller electrons  scattered at 30$^0$ with respect to the beam. This simple
device provided the luminosity  monitor: its response was independent of
{\em a)\/} luminosity fluctuations, {\em b)\/} energy of the electron 
beam and {\em c)\/} low energy electromagnetic background close to the
beam line which was measured to be less than 1\%\ of M\o ller events.

\section{Data analysis and radiative corrections}

For each run the electron yield was obtained as a function of the kinetic
energy ranging from  detection threshold ($\sim 100$ MeV) up to the tail
of QE peak. In order to obtain good statistics, the spectra
corresponding to different runs at the same kinematics were summed. In
this  analysis the following procedure was adopted:

\begin{enumerate}

\item the reproducibility of energy calibration of each spectrum was
checked within 1\%\ by determining the centroid of the proton line in
the plot of energies released in the preshower and in the calorimeter
[15];

\item all the events above threshold giving a non-zero signal in
the Cherenkov counter were then normalized to the luminosity of each run
dividing it by the number of detected  M\o ller electrons. The plot of
fig. 1 shows the result of this procedure for the spectra  collected at
880 MeV beam energy and 32$^0$ scattering angle: the fluctuations are
purely statistical and there is no evidence of a dependence on the
injected current and the jet density fluctuations;

\item data relative to each partial run were summed; then, counts were
binned in 25 MeV  energy intervals which correspond to the absolute FWHM
energy resolution for 1.5 GeV  electrons. The absolute value of the cross
section was finally obtained using the efficiency and solid angle of the
M\o ller detector as well as the M\o ller cross section. 

\end{enumerate}

Radiative corrections were calculated with a computer code [16] and
subtracted from  the spectra. In our case $t^2$ effects, $t$ being the
target thickness, were completely negligible  and corrections were
therefore applied to account for elastic radiation tail, multiple soft 
photon emission and continuum. The radiation tail for the elastic peak
was subtracted  avoiding the peaking approximation and using the exact
formula given in ref. [17] and the  measured elastic form factors for
$^{12}$C and $^{16}$O [18]. The calculation of the elastic tail was 
checked by a direct comparison to already published results [19].
Contribution of this effect,  as can be seen in fig. 2, is present in the
high energy transfer region only; in the same figure  the other
corrections discussed later on are also reported. Following ref. [17]
multiple soft  photon emission was accounted for, whereas for the
continuum contribution two different  approaches were applied for the
hard photon emission before or after the inelastic nuclear  scattering.
In the first case radiatively corrected data at lower incident energies
as well as  lower momentum transfer were needed. For this purpose we used
an interpolation of our  data when available, or the result of a
phenomenological model [20] which describes the  nuclear response
function with reasonable accuracy in the region of interest. The
radiation of  hard photons after the inelastic nuclear scattering needed,
on the other hand, an {\em ad hoc\/} evaluation due to the presence of a
non magnetic apparatus [16]. Since the angle between final electron and
radiated hard photon was well inside the angular acceptance, the total 
energy released in the detector was still the same as if the scattered
electron had not radiated. Therefore, this correction was applied at the
actual energy transfer and produced a smooth few percent reduction of
the cross section. In fig. 2 the spectrum of 880 MeV electrons  scattered
from $^{16}$O at 32$^0$ is reported at different steps of the analysis:
the small difference between raw and radiatively corrected data in the
QE peak is due to a partial compensation  between multiple soft photon
radiation process and the term describing the hard photon  emission.

As far as pair production contamination is concerned, in the present
measurement no experimental subtraction of this contribution was
possible because of the absence of a magnetic field. Terms depending
on $t^2$ were, however, completely negligible, whereas the contribution
linearly depending on $t$ was calculated following a well established
procedure [21]  which indicate no presence of such a background in the QE
region as shown in fig. 2. At higher energy transfer the pair production
contamination becomes sizable and it was found to be in agreement with
previously measured pair production spectra [1, 22]. This  background was
therefore subtracted from the data up to transferred energies where the
sum of both radiative and pair production corrections was below 40\%\ of
the measured cross  section. At very low energy transfer, contributions
to the cross section from elastic scattering  off nucleus or inelastic
transitions to bound or quasi-bound excited states are possible; 
however, in our three-momentum transfer range such contributions were
completely negligible [23].

Even though, due to the low luminosity, both electronic dead time and
pulses pile-up were negligible, some efficiency loss on the four-fold
coincidence of the electron detector could not be excluded. Moreover,
our target was not able to produce a hydrogen jet of reasonable density;
therefore a comparison with the parametrization of all the previous 
$\mbox{H}(e,e')X$ measurement in order to check absolute normalization was
impossible. Therefore, to check our apparatus performances and data
analysis we carried out some of our measurements in the same kinematical
conditions of published data [19] on $^{16}$O, namely 540  and 730 MeV
beam energy at 37.1$^0$ scattering angle. In both cases we found that a
factor 1.19  had to be applied to our data in order to reproduce the
spectra of ref. [19]: in fig.~3 the  comparison of our measurement and
those of ref. [19] performed at 540~MeV and 37.1$^0$  shows good
agreement between the two data sets; we found an analogous agreement at 
730 MeV beam energy. The good quality of data of ref. [19] and the
complete overlap of their and our cross sections give us confidence both
in our data analysis and in the  normalization procedure. The stability
of the apparatus was checked performing at the beginning and at the end
of each run (typically 7 days) the above mentioned normalization 
measurements: no variations exceeding statistical uncertainties were
found.

The total systematic error of 4.5\%\ in the QE peak was obtained from the
quadratic sum of the accuracy of radiative corrections (3\%),
calibration procedure as deduced from ref. [19] (3\%), electron beam
energy (1\%) and uncertainty on the solid angle of the detector placed
at  different angles (1\%). A value lower than 6\%\ was instead evaluated
in the D$_{13}$(1520) resonance region were both pair production
subtraction and radiative corrections become sizable.

\section{Results and comparison with theoretical predictions}

\subsection{Quasi-elastic peak}

In this section the experimental results obtained for the radiatively
corrected cross sections will be shown and compared with the theoretical
predictions for the QE region described in ref. [24]. In order to
clarify the role played by FSI, the results corresponding to the IA and 
those including an estimation of FSI will be considered. In this approach
the cross section for inclusive process $A(e,e')X$ is written in
the following form:
 \be
	   \frac{d^2\sigma}{dE_{e'}d\Omega_{e'}} = \sigma_0 + \sigma_1
 \ee
where the contributions from different final nuclear states have been
explicitly separated out, namely $\sigma_0$  describes the transition
to ground and one-hole states of the $(A-1)$-nucleon system and
$\sigma_1$ thetransition to more complex highly excited configurations.
As it is known, within the IA, the evaluation of the inclusive cross
section requires the knowledge of the nucleon spectral function     
$P(k,E)$, which represents the joint probability to find in a nucleus  a
nucleon with momentum $k\equiv |\mbox{\bf k}|$ and removal energy
$E$. In presence of ground state NN  correlations $P(k,E)$ can be
written as $P(k,E) = P_0(k,E) + P_1(k,E)$, where the indexes 0 and 1 
have the same meaning as in eq. (1), i.e. $P_0$  includes ground and
one-hole states of the $(A-1)$-nucleon system and $P_1$ more complex
configurations (mainly 1p-2h states) which  arise from 2p-2h excitations
generated in the target ground state by NN correlations. Whereas $P_0$ 
depends on the nucleon momentum distribution of the single particle
states, for the correlated part $P_1$ we will make use of the model
developed in refs. [25, 26], which involves the basic two-nucleon
configurations generating the high momentum and high  removal energy
behaviour of the spectral function. Finally, besides the two spectral 
functions $P_0$ and $P_1$, the cross section of eq. (1) depends on 
$\sigma_{eN}$, describing the scattering of the electron by an off-shell
nucleon as computed in [27].

Comparison of our results in the QE region with the described IA approach
is reported in fig. 4 by the dashed line. The experimental data of
$^{16}$O were collected at 32$^0$ scattering  angle and different beam
energies. The inclusion of transitions to highly excited  configurations
accounted for by $P_1$ is not sufficient to explain the experimental
cross section. At the QE peak the calculations largely overestimate the
measured strength especially for the data at 700 MeV beam energy,
whereas an opposite effect is observed in all the collected  spectra in
the low energy transfer region corresponding to $x \geq 1.5$.

In order to account for such differences, both single and two nucleon
rescattering have  been included, following the approach of ref. [24].
This approach treats consistently the  effects of NN correlations in both
initial and final nuclear states: in particular, a locally  correlated NN
pair with its centre of mass apart from the spectator $(A-2)$ nucleus
is  considered [25, 26] and the two emitted nucleons are allowed to
rescatter elastically in the final states. It should be pointed out that
the approach of ref. [24] has been positively checked against SLAC
data [28] both for light and complex nuclei at $x > 1$ and high 
momentum transfer ($Q^2>1$ (GeV/c)$^2$). In the present paper the same
approach is applied to  the calculation for lower values of $Q^2$ 
($0.1 < Q^2 < 0.6$ (GeV/c)$^2$) and the result, together with our
experimental data, is represented in fig. 4. The inclusion of such FSI is
sizable: from the QE peak the strength is correctly moved to both low
and high energy transfer regions greatly improving the agreement with
the experimental data and extending the $Q^2$ interval where this
theoretical approach can describe the data.

\subsection{The resonance region}

At energy transfer higher than the QE peak, the inclusive cross section
is dominated by the pion-nucleon resonance, the $\Delta$(1232) and, for
our highest beam energy, by the less prominent N(1520). This is clearly
evident from fig. 5, where some of our inclusive spectra of $^{16}$O are
reported up to $W\approx 1500$ MeV. In the following analysis the
inclusive cross sections on $^{12}$C from ref. [4] were also used. The
data have been divided by the nuclear mass $A$ but the slightly
different kinematics (scattering angle and $Q^2$) do not allow a
direct comparison of the measured strength for the two different
nuclei. However, when the  normalization to the virtual photon flux
was also performed, a response well within the total  uncertainty was
obtained as shown in fig.~6 for two measurements with similar $Q^2$ in
the $\Delta$(1232) peak.

Different approaches have been attempted in order to describe the
excitation mechanism  in this region. An evaluation of the pion
electroproduction cross section on single nucleon taking into account
resonant, non-resonant (Born) terms and final state interactions [29]
was  extended to finite nuclei [30] including the two-body
$\gamma \mbox{NN} \rightarrow \mbox{NN}$ mechanism which is assumed to
be dominated by the coupling to the $\pi$NN intermediate state. In order
to reproduce the experimental data, the medium effects on the
propagation of $\Delta$(1232) were included by changing the
$\Delta$(1232) self-energy from its value in the free space. The
$\Delta$-hole approach [31] carefully describes the resonant part of the
reaction mechanism: this model, successfully applied to the description
of pion- and photon-induced nuclear reactions in the resonance region,
includes dynamicaleffects like Pauli blocking and pion multiple
scattering. These calculations have been compared to the experimental
data on light nuclei at a relatively  low momentum ($Q^2 \leq 0.15$
(GeV/c)$^2$) in the $\Delta$(1232) peak: in both cases the predicted
cross  section turns out to be lower than the data from the dip to the
$\Delta$(1232) peak by 15\%. Our data were, instead, collected in a
higher momentum transfer region ($0.1 \leq Q^2 \leq 0.5$ (GeV/c)$^2$)
where the impulse approximation is expected to dominate: in order to find 
possible medium modifications to the single nucleon strength, we compared
the experimental data to a simple calculation where only Fermi motion
effects were taken into consideration. The main steps of our analysis
are here summarized.

Starting from the well known Brasse parametrization of experimental
electron  scattering cross section on free proton [32], we fitted more
recent H$(e,e')$  data [4], measured in kinematical conditions very
similar to ours, by a small tuning of the original parameters of the
fit. Then, following the prescription of ref. [33], the result of our 
parametrization was folded on the nucleon momentum distribution given in
ref. [25] to  obtain the inelastic structure functions in the nucleus.
The same distribution was assumed  both for our data in $^{16}$O and for
$^{12}$C data of ref. [4] and no difference in the virtual photon 
absorption on proton $\sigma_p$ and neutron $\sigma_n$ was taken into
account. The result of this calculation is reported in fig. 5 together
with both our measurements in $^{16}$O and ref. [4] data in $^{12}$C for
different beam energies and scattering angles. The QE peak contribution, 
evaluated as described in the previous section, was directly summed to
the inelastic part and  the result compared to the data: this comparison
immediately shows that the resonance  structure has a broader shape than
the calculation, whereas a damping of the inelastic  strength with
respect to the single nucleon seems to be required at the higher $Q^2$
values. This is even more evident for $^{12}$C data (fig. 5{\em c\/} and
{\em d\/}) measured with less statistical uncertainty. It is important
to note that these data have been taken in exactly the same kinematical
conditions  of the H$(e,e')$ reaction we used to evaluate the single
particle inelastic strength.

This result is consistent with the known fact that the $\Delta$(1232)
resonance in nuclei is  broadened by an additional width beyond the
natural decay width and Fermi motion [2]: to  single out this nuclear
effect on the resonance cross section it was therefore necessary to 
develop first a suitable model for $\Delta$(1232) excitation on free
nucleon. We assumed a  relativistic approach where the correct treatment
of the kinematics ensures that the $\gamma N\Delta$  transition form
factor is evaluated at the proper $Q^2$. Thus we used the $\Delta$
propagator of the  Rarita-Schwinger theory, whereas the Peccei
Lagrangian [8, 34, 35] provided the $\gamma N \Delta$ vertex. The 
$\gamma N \Delta$ coupling constant is usually chosen to reproduce the
resonant channel M1$^+$ of $\gamma \mbox{N} \rightarrow \pi N$     
reaction. Since, however, the Peccei Lagrangian provides an abnormally
high contribution for the Coulomb multipole of about 15\%, we simply
rescaled the peak of the same amount to account for M1$^+$ transitions
only. Moreover, $\Delta$(1232) cannot be considered as stable. We thus
added in the denominator of the propagator its width in the vacuum 
(dependent on energy and momentum to account, for instance, for threshold
effects [36]). Its  strong coupling constant and form factor were finally
fixed by the elastic $\pi$N cross section.

As far as higher energies are concerned, at $W \approx 1500$ MeV      
two resonances are observed in the nucleon: N(1520) and N(1535) with a
full width similar to $\Delta$(1232). Since this 15 MeV mass difference
is far below the energy resolution of our apparatus, we considered only
the N(1520) resonance which was assumed to have the same dependence of 
the cross section on the energy transfer as the $\Delta$(1232). The
strength was determined from H$(e,e')$ data as reported, in one
example, in fig. 7; the continuous curve is the result of the  modified
Brasse fit we used to determine the inelastic part whereas the dashed
lines represent the separated contributions to the total strength: our
calculated $\Delta$(1232) excitation curve, the  N(1520) peak---whose
amplitude was kept as a free parameter---and a phenomenological 
non-resonant background. For the latter the expression
 \be
   	B(W) = \alpha(W - W_0)^{\beta}
 \ee
was assumed where $W_0$ is the pion electroproduction threshold and
$\alpha$, $\beta$ are free parameters. In all the spectra we have
assumed the parameter $\beta$ to be linearly dependent on the four 
momentum value at the pion electroproduction threshold $Q^2_{th}$:
 \be
   	\beta = 0.70 Q^2_{th}- 0.10
 \ee
whereas the values of $\alpha$ were sensitive to the specific kinematical
conditions.

In order to extend this analysis to the nucleus, the two resonances and
the continuous background were separately folded with the momentum
distribution and shifted by the average nucleon binding energy. In view
to highlight possible medium modification to the quasi-free picture, the
model amplitude and width have been therefore adjusted to fit the
observed data in the high energy transfer region both for $^{16}$O and
$^{12}$C. As a first step we broadened both  resonances by an additional
width (in quadrature) up to a maximum value corresponding to  twice the
experimental value reported in ref. [2]. Result of this procedure is
reported for our and ref. [4] data in fig.8. Here the two broadened
resonances and the continuous  background are separately plotted, while
the total sum including QE peak calculation is  represented by the
continuous curve: whereas for the low beam energy data this widening 
accounts for the observed strength, at higher energies (higher $Q^2$) an
excess of strength  which can not be reduced, at least for $^{12}$C data,
by a further spreading out of resonances is nevertheless present. This
could be an indication that, accordingly to the real photon case [37],  a
proper suppression factor must be simultaneously applied to original
resonance  amplitude. In fig. 9 the thin curve corresponds to the result
of this procedure: it is now evident that a proper suppression factor to
the resonant strength can account for the experimental data.

However, the result of this analysis is not unambiguous, especially in
N(1520) region where the data are limited to $W\leq 1500$ MeV. In the
same figure we have, in fact, also reported the computed strength when
both the broadened resonances and the remaining inelastic part are
damped by the same factor, obtaining a very similar result. The limited 
discrepancy observed in the high energy transfer side of the QE peak
might be ascribed to the folding with momentum distribution instead of
spectral function in the inelastic part  calculation.

The results of the analysis are plotted in fig. 10 and summarized in
table I: here $\sigma$ represents the extra width applied to the
resonances, $a$ and $b$ are the cross section suppression factors for
the $\Delta$(1232) and N(1520) resonances respectively, $c$, instead, 
represents the overall suppression factor when both the resonant and the
non-resonant part are simultaneously changed. This analysis shows that
the extra width $\sigma$ necessary to reproduce the shape of
$\Delta$(1232) peak is quite independent from the kinematics, whereas
the suppression factors seem to be stronger as $Q^2$ increases. If,
for the $\Delta$(1232) resonance, this  factor approaches to unity at low
$Q^2$, for the N(1520) resonance a damping is always  observed giving
some sort of continuity with respect to the real photon case. The 
consistency of the extra width and suppression factor for $^{12}$C and
$^{16}$O data suggests a similar  behaviour of the response function of
the two nuclei in the analyzed momentum range. Even if the quoted
errors are statistical only, corresponding to a unit change on $\chi^2$,
the inclusion  of the systematic error (6\%\ for our data and 3\%\ for
the ref. [4] ones) does not seem to significantly change the results of
this analysis.

A major source of indetermination could, instead, be introduced by a
possible difference between proton $\sigma_p$ and neutron $\sigma_n$ 
absorption strengths. A parametrization which takes into account this
difference [38] gives, in fact, $\sigma_n/\sigma_p \cong 0.76$         
in the kinematical range of our and  ref. [4] data. Using this
parametrization, the suppression factors found in the previous  analysis
are reduced. This fit, however, is the result of the extrapolation of
proton and  deuteron data measured at higher $Q^2$ ($Q^2>1$
(GeV/c)$^2$) and its accuracy in our kinematical range might be
questionable. In order to give a more quantitative evaluation of the
proton-neutron difference, the following considerations can be done. Due
to the isospin structure of the interaction, the proton and neutron
helicity amplitudes relative to the $\Delta$(1232) transition  are equal
and any difference of $\sigma_n$ from $\sigma_p$ comes therefore from
the Born terms only. However, both our fitting procedure on the 
H$(e,e')$ data and the result of the complete  calculation of
refs. [29, 30] indicate that in the $\Delta$(1232) region and in our
kinematical conditions this non-resonant contribution accounts for less
than 50\%\ of the total strength. The overall damping observed in our
and ref. [4] data at  $Q^2\approx 0.5$ (GeV/c)$^2$ is 25\%\
and the difference of $\sigma_n$ from $\sigma_p$  which could account
for this suppression is therefore $\sigma_n/\sigma_p = 0.5$. This is a
too small value, which would imply that the Born terms for the neutron
are completely negligible in contradiction to the basic mechanisms for
the virtual  photoabsorption off the nucleon, where the pion in flight
contribution is sizable both for proton and neutron.

Finally, as an alternative hypothesis, we increased the non resonant
background contribution in order to simulate a possible multiparticle
emission which could be present in our kinematical conditions. In fact,
it is well known that, as one proceeds to investigate the  nuclear
response in the inelastic region, the one particle-hole frame is no
longer adequate, the inclusion of two particle-two hole excitations
becoming more and more important [39]. We analyzed, therefore, the
$^{16}$O and $^{12}$C measurements at 1500 MeV where the continuous 
background to the resonant contribution ratio is maximized. For an
enhancement of the background up to 15\%\ we could still find a good fit
to the data (comparable $\chi^2$ as fig. 10  results) provided that a
further simultaneous damping was applied to the resonances, still 
supporting our previous conclusions.

\section{Conclusions}

We have measured the $(e,e')$ inclusive cross section on a pure
$^{16}$O jet target with a shower calorimeter on a wide range of energy
and momentum transfer. The results in the QE region  have been compared
to a quite extensive calculation. Along with a realistic spectral
function which contains a correlated part, the inclusion of FSI is
necessary for a correct description of the data particularly at low
momentum transfer. In the inelastic region, besides the Fermi  motion
effect, a widening and a damping of the resonances is observed in our
upper limit of $Q^2$ ($0.2 \leq Q^2 \leq 0.5$ (GeV/c)$^2$) with respect
to the single nucleon strength. In order to understand this effect,
realistic calculations including medium dependent modifications as  well
as the difference between proton and neutron absorption should be
developed.

\newpage

\section*{Table legend}

{\bf Tab. 1.} Results of our analysis in inelastic region: $\sigma$ is
the extra widening of resonances, $a$ and $b$ are the suppression
factors of $\Delta$(1232) and N(1520) resonances respectively, $c$ is
the  overall suppression factor applied to the whole inelastic
contribution.

\newpage

\section*{Figure legends}

\hspace{\parindent} {\bf Fig. 1.} Ratio $R$ of the detected electrons to
the luminosity for each run. The collected data correspond to $E_e =
880$ MeV electron beam energy and $\theta_{e'} = 32^0$  scattering angle.

{\bf Fig. 2.} Corrections applied to the spectrum at 880 MeV beam energy
and 32$^0$ scattering angle. The experimental data (open triangles) and
the radiatively corrected data (black square) are shown together with
the elastic radiative tail (full curve), the hard photon emission before 
(dot-dashed) and after (dotted) nuclear scattering and the pair production
contribution (dashed). The soft photon emission correction is not
reported.

{\bf Fig. 3.} Comparison of $^{16}$O$(e,e')$ inclusive cross section
measured in the present  experiment (open squares) with the results of
ref. [19] (full dots) at 540 MeV electron beam  energy and 37.1$^0$
scattering angle.

{\bf Fig. 4.} Inclusive cross section measured for the process    
$^{16}$O$(e,e')$ versus the energy  transfer $\omega$ in three
kinematical conditions corresponding to the electron scattering angle 
$\theta_{e'} = 32^0$ and incident energy {\em a)\/} $E_e = 700$, {\em
b)\/} 1080 and {\em c)\/} 1200 MeV. The dashed lines represent the
theoretical predictions obtained within the IA, the solid lines include
the effects of the FSI of the knocked-out nucleon with the residual
nuclear system evaluated according to ref. [26]. On the top of each plot
the value of the Bjorken scaling variable $x$ is also reported.

{\bf Fig. 5.} Experimental inclusive cross section in {\em a,b)\/} 
$^{16}$O (this experiment) and {\em c,d)\/} $^{12}$C [4] in different
kinematical conditions. The data are compared with the calculation (full
curve),  which includes the QE contribution (dotted) as well as the
inelastic strength directly deduced  from the free proton (long dashed).

{\bf Fig. 6.} The inclusive cross section per nucleon normalized to the
virtual photon flux vs. the invariant mass $W$. The open squares and
full circles are the data for $^{16}$O and $^{12}$C; the $Q^2$ value at
the $\Delta$(1232) peak is 0.25 and 0.26 (GeV/c)$^2$, respectively.

{\bf Fig. 7.} Example of the H$(e,e')$ cross section data [4] fitted with
our version of the Brasse [32] parametrization (full curve); the dashed
lines represent the resonant and non-resonant contributions to the cross
section as determined in our analysis.

{\bf Fig. 8.} Inclusive  $(e,e')$ cross section in {\em a,b)\/} $^{16}$O
(this experiment) and {\em c,d)\/} $^{12}$C (ref. [4]). The different
contributions to the inelastic cross section are shown: the non-resonant 
background (dotted), the D(1232) and the N(1520) resonances (dashed
curves). At this step of the analysis the resonance cross-section was
broadened by the additional width reported in table I. The full curve
represents the sum of this parametrization and the QE contribution.

{\bf Fig. 9.} Final result of our analysis (QE + inelastic), where an
extra widening of the  resonant part and a suppression factor is applied
to the $\Delta$(1232) and N(1520) resonances only (thin curve) or to all
the inelastic contribution (thick curve). The corresponding suppression
factors are reported in table I.

{\bf Fig. 10.} The suppression factors as deduced from our analysis for 
$^{16}$O (full squares) and $^{12}$C (open circles) as a function of
$Q^2$ at fixed invariant mass $W$. The result of the analysis [37]  at
the photon point is also reported (open triangles).

\newpage

\section*{Table 1}

\begin{tabular}{|c|c|c|c|c|c|} \hline
		&$E_0$, $\theta$ &$\sigma$	&$a$			&$b$		&$c$\\ 
		&(MeV, deg)	&(MeV)		&			&		& \\ \hline\hline
$^{16}$O	&700, 32	&$50 \pm 15$	&$1.00 \pm 0.05$	&---		&$1.00 \pm 0.05$
\\ \cline{2-6}
		&880, 32	&$50 \pm 15$	&$0.90 \pm 0.05$	&---		&$0.90 \pm 0.05$
\\ \cline{2-6}
		&1080, 32	&$50 \pm 15$	&$0.95 \pm 0.07$	&---		&$0.95 \pm 0.07$
\\ \cline{2-6}
		&1200, 32	&$75 \pm 15$	&$1.00 \pm 0.07$	&$0.8 \pm 0.1$	&$0.95 \pm 0.07$ 
\\ \cline{2-6}
		&1500, 32	&$80 \pm 25$	&$0.50 \pm 0.08$	&$0.4 \pm 0.2$	&$0.75 \pm 0.07$ 
\\ \hline $^{12}$C	&960, 37.5	&$50 \pm 10$	&$0.85 \pm 0.04$	&---		&$0.90
\pm 0.04$ \\ \cline{2-6}
		&1100, 37.5	&$75 \pm 10$	&$0.85 \pm 0.04$	&---		&$0.90 \pm 0.04$
\\ \cline{2-6}
		&1200, 37.5	&$60 \pm 10$	&$0.80 \pm 0.04$	&$0.8 \pm 0.1$	&$0.90 \pm
0.04$ \\ \cline{2-6}
		&1500, 37.5	&$50 \pm 10$	&$0.60 \pm 0.04$	&$0.4 \pm 0.2$	&$0.80 \pm
0.04$ \\ \hline
\end{tabular}

\newpage

\begin{figure}

\epsfig{file=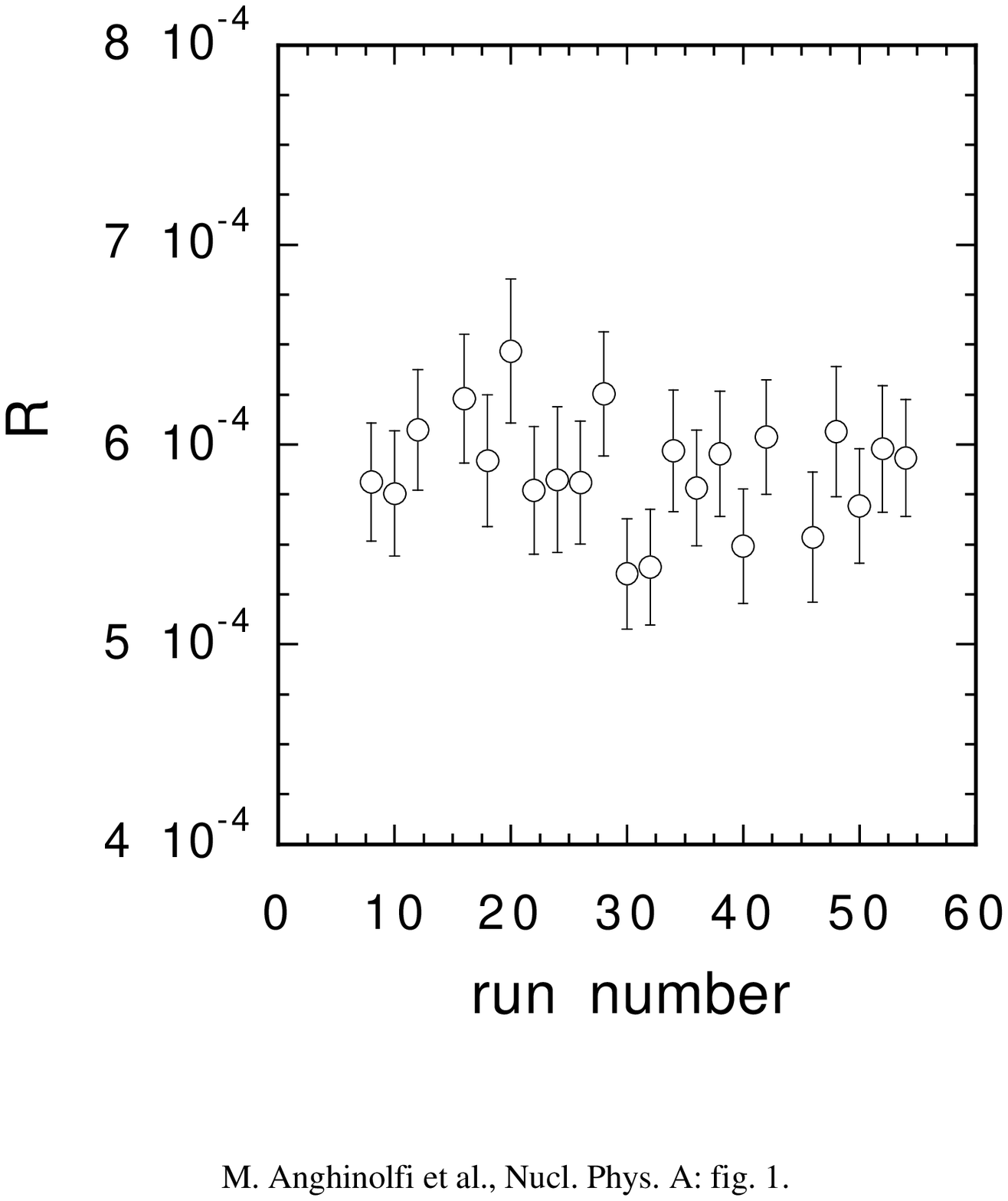}

\end{figure}

\newpage

\begin{figure}

\epsfig{file=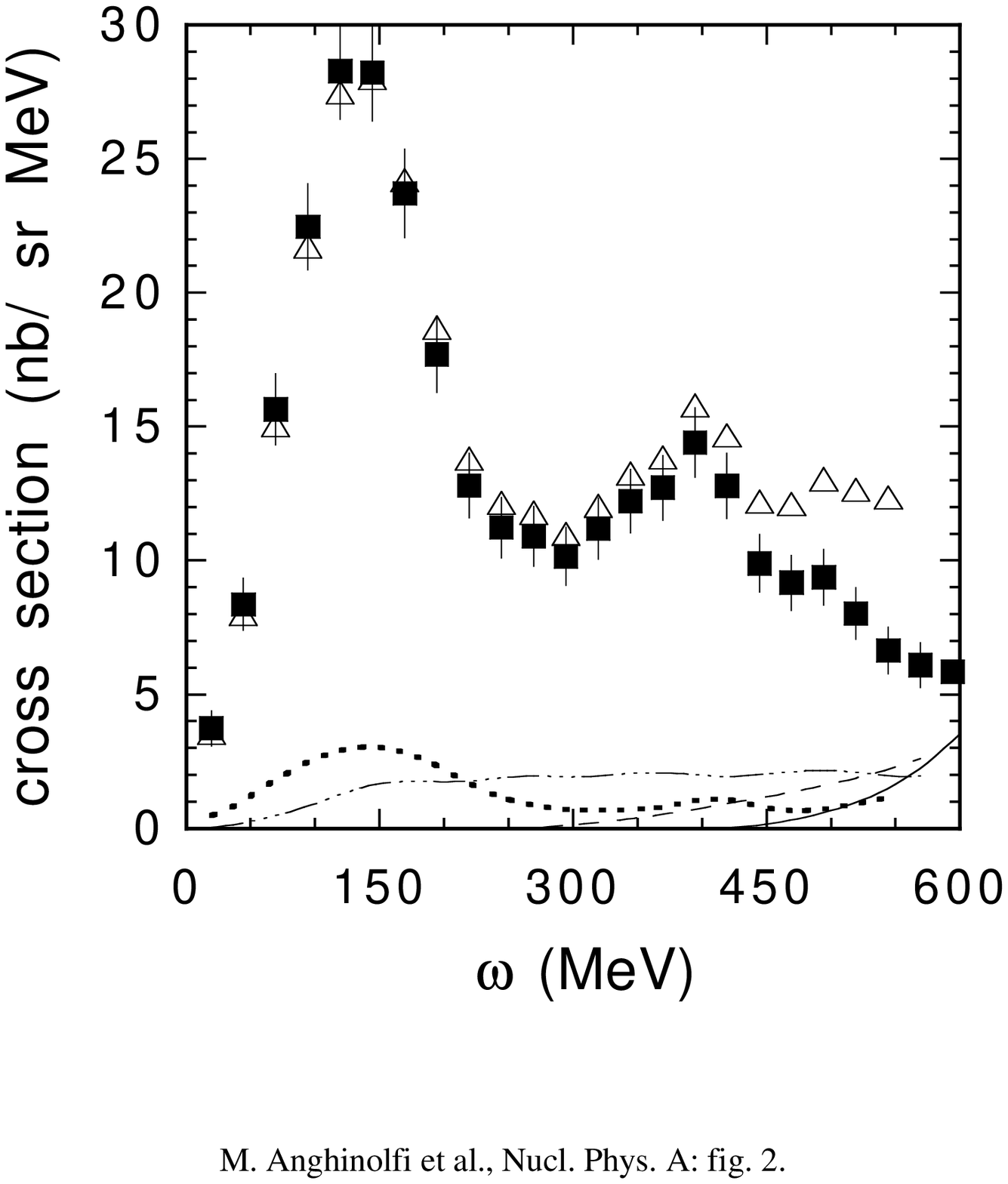}

\end{figure}

\newpage

\begin{figure}

\epsfig{file=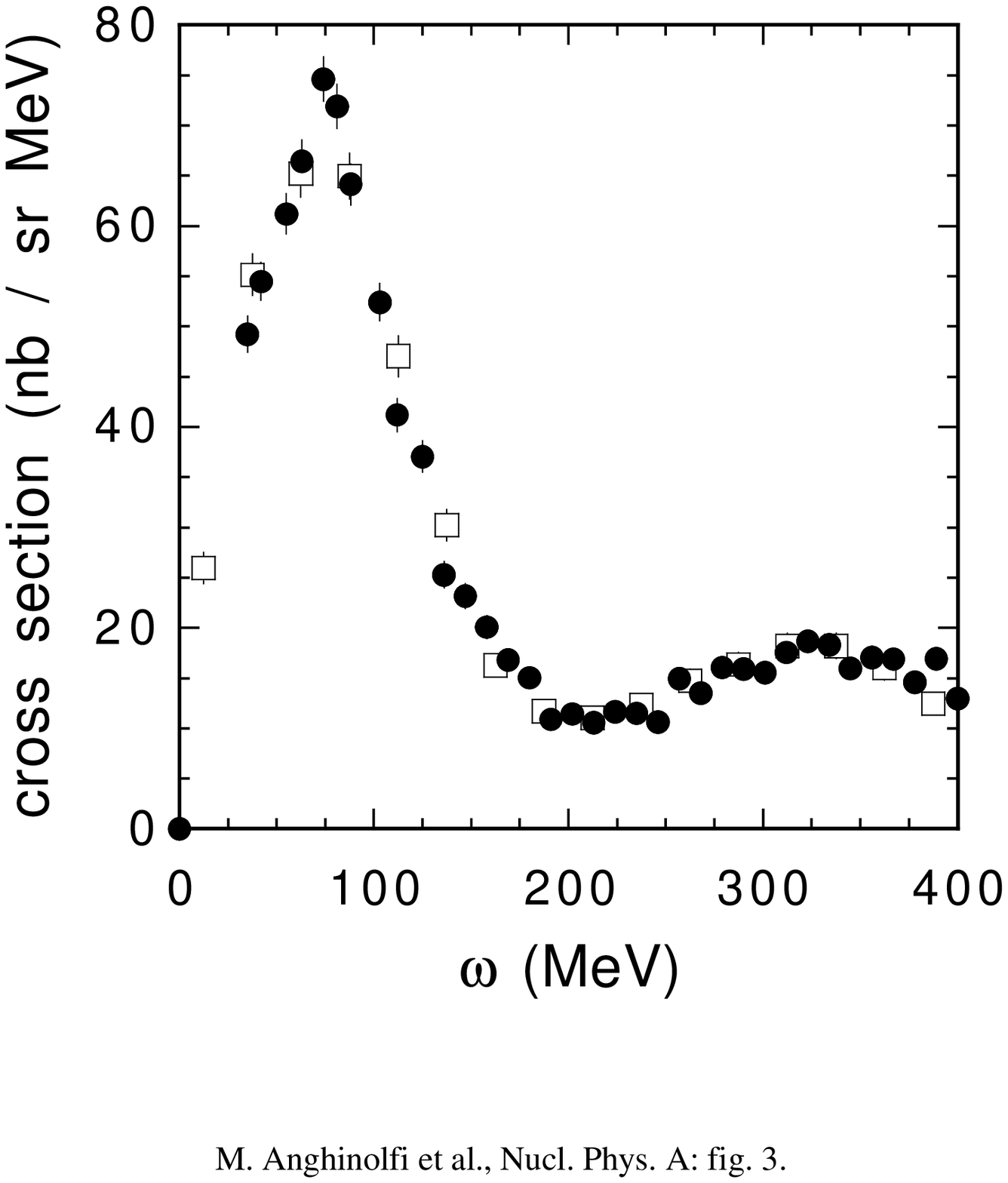}

\end{figure}

\newpage

\begin{figure}

\epsfig{file=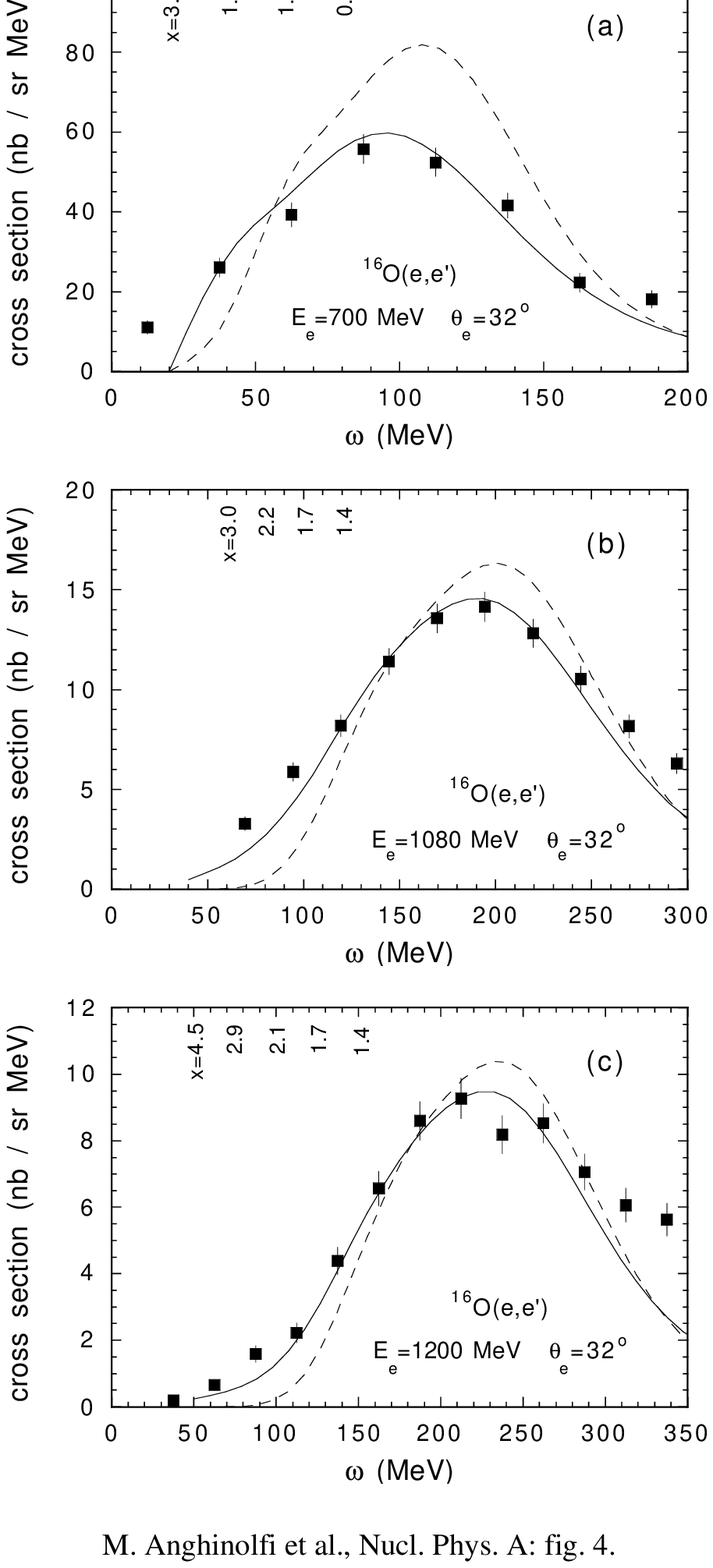}

\end{figure}

\newpage

\begin{figure}

\epsfig{file=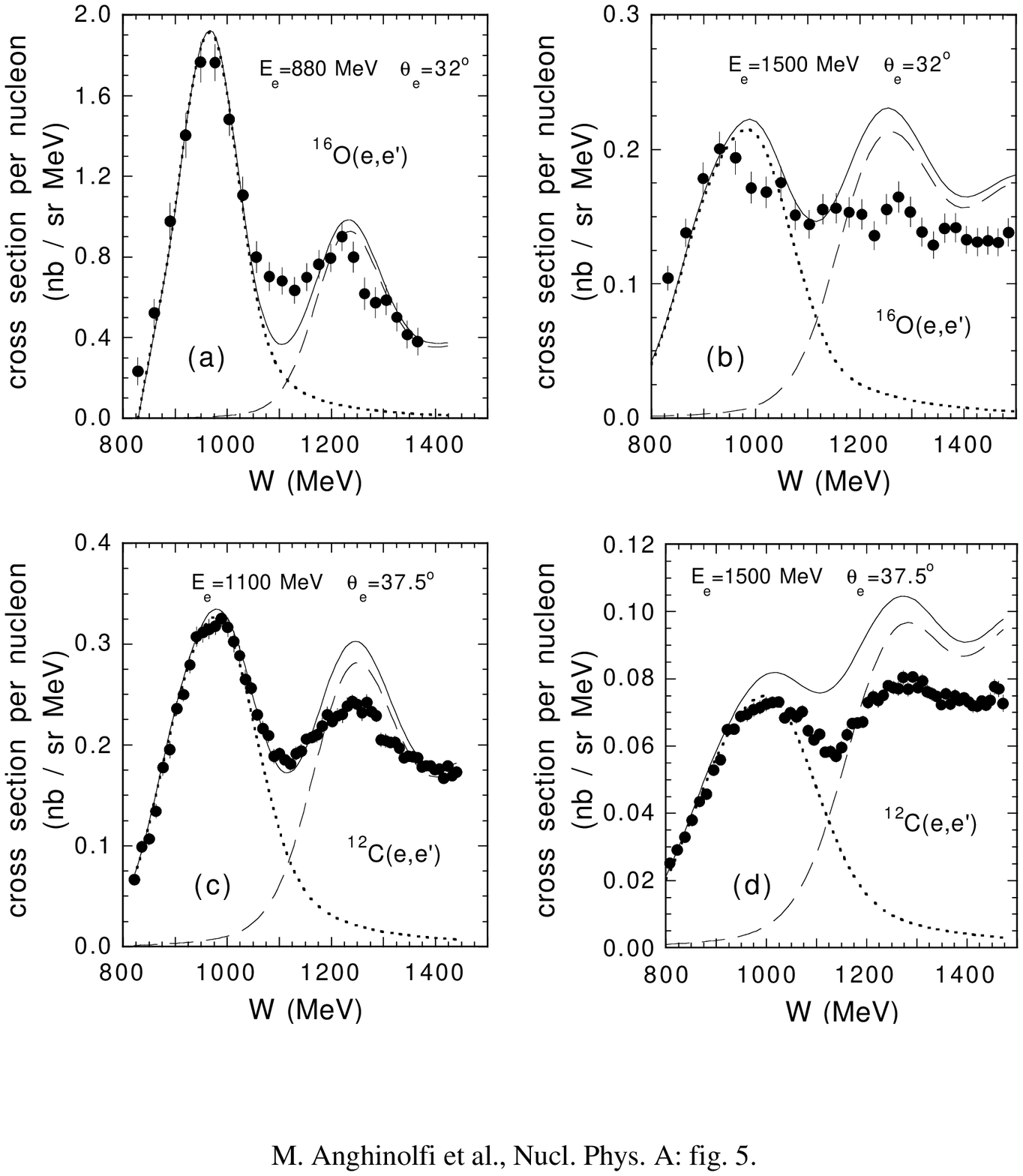}

\end{figure}

\newpage

\begin{figure}

\epsfig{file=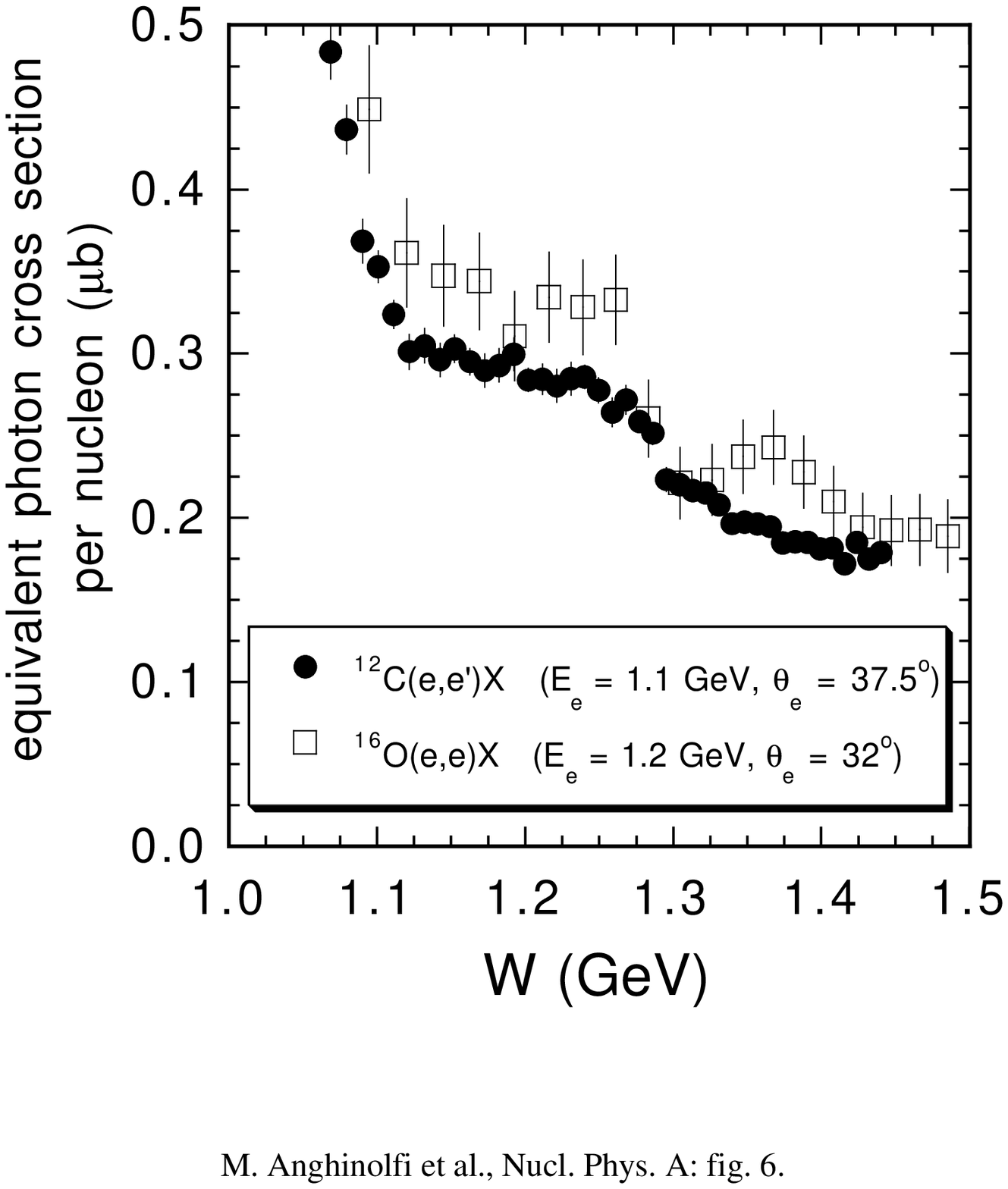}

\end{figure}

\newpage

\begin{figure}

\epsfig{file=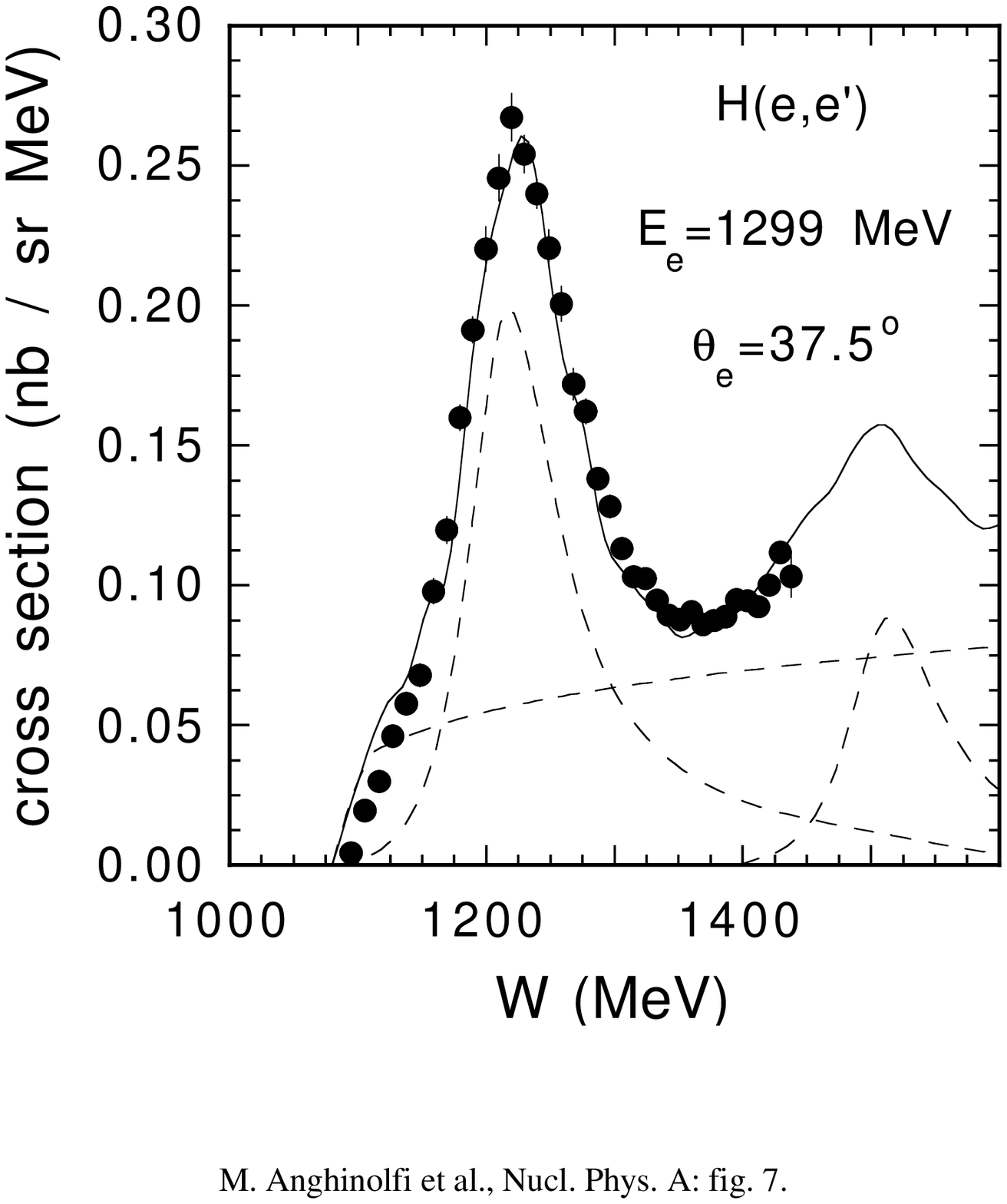}

\end{figure}

\newpage

\begin{figure}

\epsfig{file=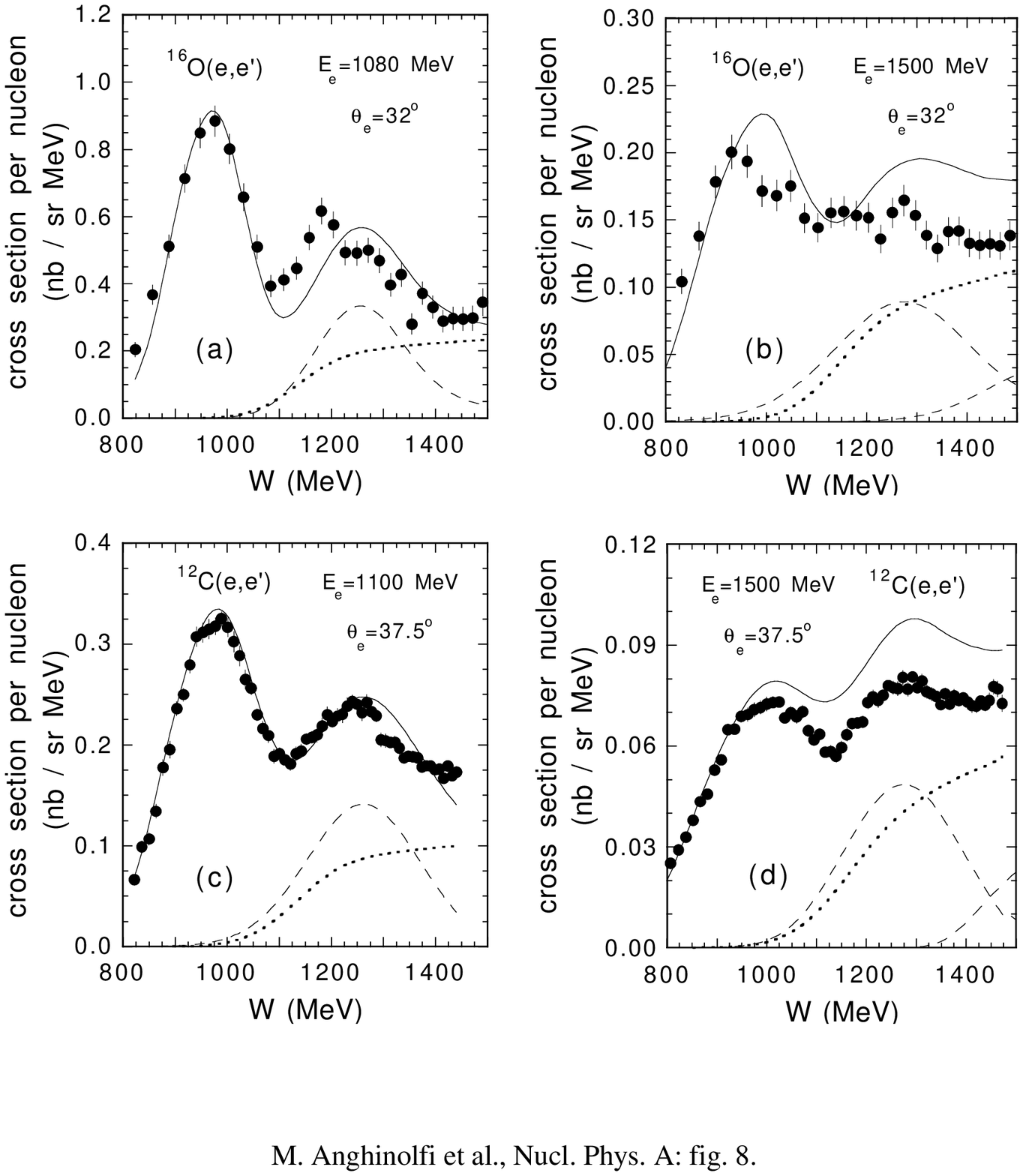}

\end{figure}

\newpage

\begin{figure}

\epsfig{file=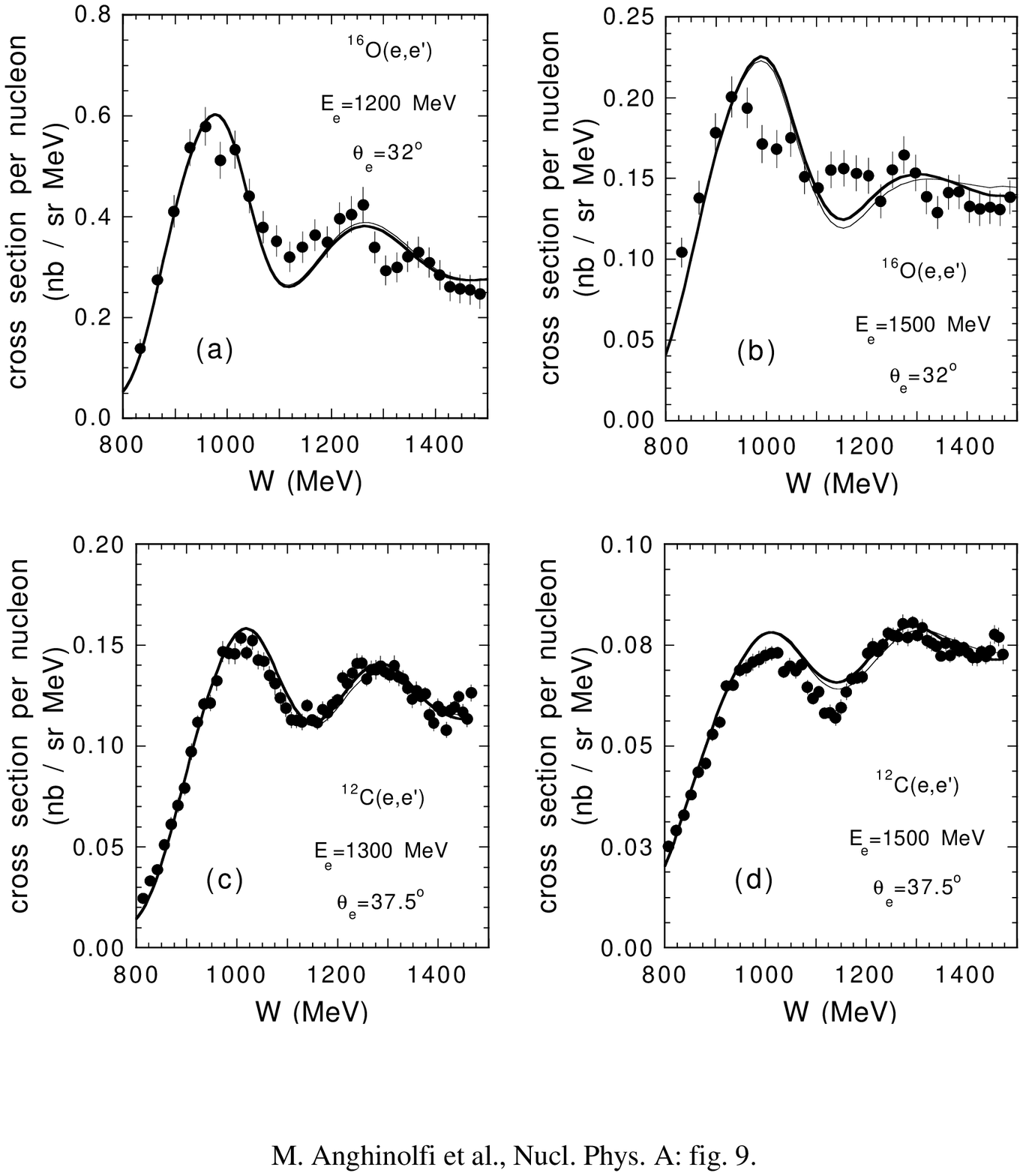}

\end{figure}

\newpage

\begin{figure}

\epsfig{file=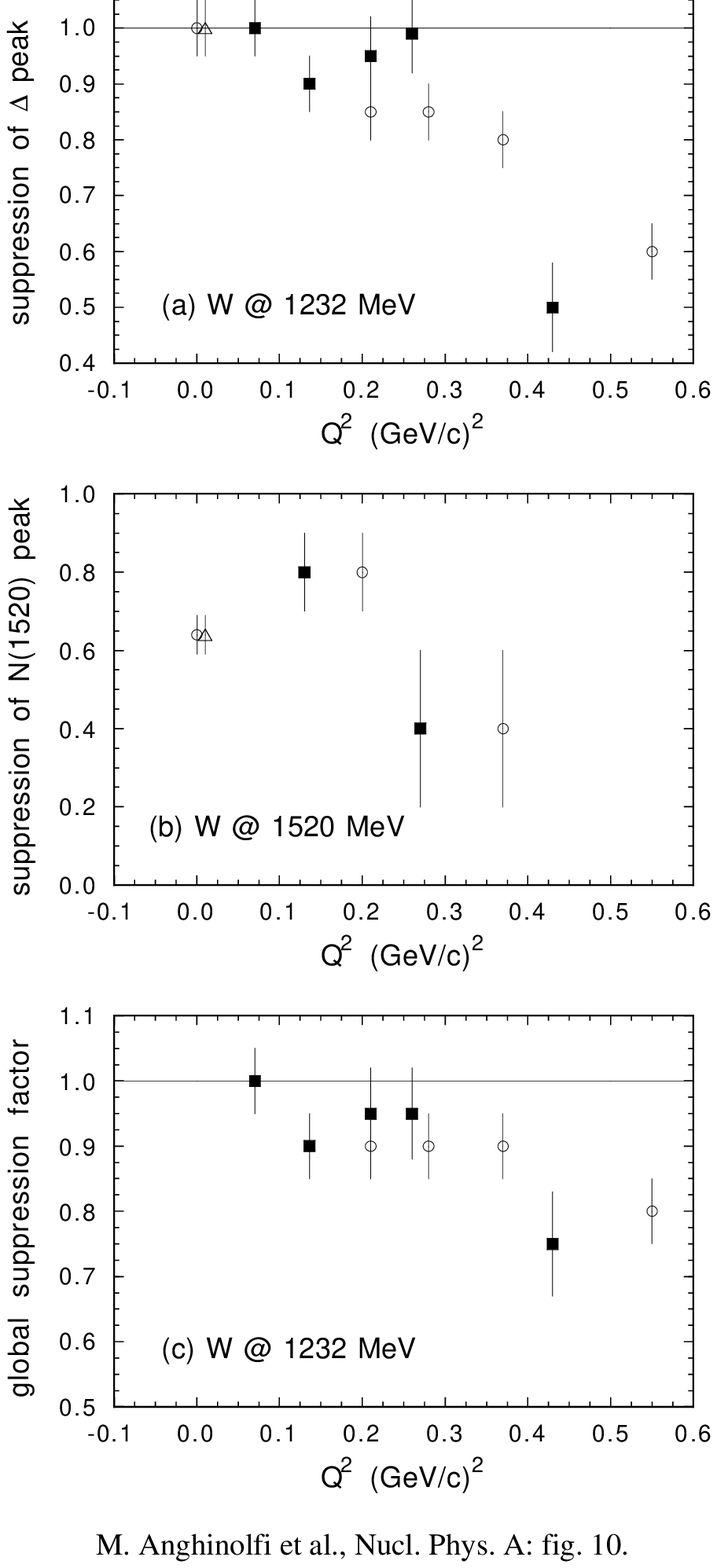}

\end{figure}

\end{document}